\documentclass[preprintnumbers,prd,onecolumn,floatfix,superscriptaddress, nofootinbib]{revtex4-2}

\usepackage{setspace}
\usepackage{graphicx}
\usepackage{subfig}
\usepackage{epsfig}
\usepackage{bm}
\usepackage{amssymb}
\usepackage{float}
\usepackage{amsmath}
\usepackage{dcolumn}
\usepackage[colorlinks]{hyperref}
\usepackage[usenames,dvipsnames]{color}
\usepackage{enumitem}

\hypersetup{ breaklinks=true, pdfstartview={FitH}, colorlinks=true, linkcolor=blue, citecolor=red, filecolor=magenta, urlcolor=blue, anchorcolor=green, linktocpage=true }

\renewcommand{\(}{\left(}

\def\doi{http://doi.org}

\def\t{\tilde}

\newcommand{\be}{\begin{equation}}
\newcommand{\ee}{\end{equation}}
\newcommand{\beano}{\begin{eqnarray*}}
\newcommand{\eeano}{\end{eqnarray*}}
\newcommand{\ba}{\begin{eqnarray}}
\newcommand{\ea}{\end{eqnarray}}

\begin{document}

\title{Bouncing cosmology in modified gravity with higher-order curvature terms}
\author{J. K. Singh}
\email{jksingh@nsut.ac.in}
\affiliation{Department of Mathematics, Netaji Subhas University of Technology, New Delhi-110 078, India}
\author{Harshna Balhara}
\email{harshnabalhara24@gmail.com}
\affiliation{Department of Mathematics, Netaji Subhas University of Technology, New Delhi-110 078, India}
\author{Kazuharu Bamba} 
\email{bamba@sss.fukushima-u.ac.jp}
\affiliation{Division of Human Support System, Faculty of Symbiotic Systems Science,\\ Fukushima University, Fukushima 960-1296, Japan}
\author{J. Jena}
\email{jjena67@rediffmail.com}
\affiliation{Department of Mathematics, Netaji Subhas University of Technology, New Delhi-110 078, India}

\begin{abstract}
\begin{singlespace}

A bouncing scenario of a flat homogeneous and isotropic universe is explored by using the reconstruction technique for the power-law parametrization of the Hubble parameter in a modified gravity theory with higher-order curvature and trace of the energy-momentum tensor terms. It is demonstrated that bouncing criteria are satisfied so that the cosmological initial singularity can be avoided. In addition, it is shown that the equation of state parameter crosses the line of the phantom divide. In the present scenario, the universe is filled with a perfect fluid around the bouncing point, in which the universe becomes highly unstable and a big bounce can be realized. Furthermore, it is found that extremal acceleration occurs at the bouncing point.

\end{singlespace}
\end{abstract}

\maketitle

\noindent
PACS numbers: {04.50.-h, 98.80.-k.}\\

Keywords: {modified gravity theory; bouncing scenario; quintom model}.

\section{Introduction}
The current accelerated expansion of the universe is confirmed by the recent cosmological observations such as Supernova type Ia (SNeIa), Baryon Acoustic Oscillation (BAO), Cosmic Microwave Background (CMB) radiation, Large Scale Structures (LSS) of the universe, Baryon Oscillation Spectroscopic Survey (BOSS) collaboration, the Planck satellite \cite{Planck:2018vyg, Planck:2018jri} and the Weak Lensing. To explain the current cosmic acceleration, the energy component with its negative pressure called ``dark energy (DE)'' is necessary. According to the Planck results \cite{Planck:2018vyg}, it is known that about 68\% of the energy fraction of the current universe is dark energy, 27\% of it is dark matter, and about 5\% of it is baryon. 

A number of candidates for dark energy have been proposed, e.g., quintessence \cite{Farooq:2011yz}, phantom \cite{Jamil:2011za, Jamil:2011sx}, k-essence \cite{Pasqua:2011gh}, and Chaplygin gas \cite{Chaplygin}. Despite the success of current dark energy cosmological models, there are still unresolved issues, which has led theorists to explore alternative theories of gravity as a possible explanation. Modified theories of gravity that mimic the late-time acceleration of the universe have been developed by altering the Einstein-Hilbert action. 
Many studies have been conducted to explore the implications of these modified gravity theories on the acceleration of the universe (for several reviews of dark energy and modified gravity theories, see, Refs. \cite{Nojiri:2010wj,Nojiri:2006ri, padmanabhan2007, Copeland2006, Durrer2008, Clifton2012, Bamba2012, Capozziello2011, Nojiri2017, Sotiriou2008, DeFelice2010, Bamba:2015uma, Cai:2015emx}). The analyses of the late-time cosmic acceleration have been explored in modified gravity theories \cite{Nojiri:2003vn, Sahni:1999gb, Tsujikawa:2007xu, Nojiri:2006gh, Sharma:2014zya,Myrzakulov:2012qp, Starobinsky:1980te, Moraes:2017zgm, Shabani:2014xvi, Alves:2016iks, Abedi:2018lkr, DAgostino:2018ngy}.

In this work, we concentrate on $\mathcal{F}(R,T)$ gravity with $R$ being the Ricci scalar and $T$ the trace of the stress energy-momentum tensor \cite{Harko:2011kv} because it is considered to be an efficient and promising theory of gravitation. The matter Lagrangian $ \mathcal{L}_m $ in $ \mathcal{F}(R,T) $ gravity is changed in relation to the space-time which is indicated by the presence of the source term. Since the expression of the source term is generated as a function of $T$, different representation of $T$ can influence the distinct sets of field equations. In $ \mathcal{F}(R,T) $ gravity, the motion of classical particles does not follow the geodesics because the covariant divergence of the stress energy-momentum tensor does not disappear. Such an extra acceleration is sufficient to realize the late-time cosmic acceleration, although the law of the energy-momentum conservation is broken. Many aspects of cosmology have been investigated in the framework of $ \mathcal{F}(R,T) $ gravity \cite{Houndjo:2011tu, Wu:2018idg, Singh:2018xjv, Shabani:2016dhj, Singh:2014kca, Singh:2022eun, Nagpal:2018uza}.


 The early universe began with a big bang but had cosmological issues such as the horizon problem, original structure problem, flatness problem, entropy problem, transplanckian problem and singularity problem. Inflationary cosmology was established \cite{Guth:1980zm, Sato:1980yn} to address these issues, where the universe expanded exponentially for a short period of time $10^{-30} $ seconds after the big bang. 
 Inflation can explain almost all of the issues raised above. However, the singularity problem still remains, and therefore the concept of the big bounce has been considered as a complementary idea to inflation. This idea incorporates the so-called cyclic model in which the first cosmological event followed the collapse of the previous universe.
  
In bouncing cosmology, the phase of the universe changes from the initial contraction one to the late-time expansion one at the bouncing point, in which the Hubble parameter $ H(t) $ changes from $H(t) < 0$ to $H(t) > 0$, and at the bouncing point, $ H(t) = 0$. In a spatially flat Friedmann-Lema\^{i}tre-Robertson-Walker (FLRW) universe, it is widely known that the Null Energy Condition (NEC) for matter must be broken in order to achieve a successful nonsingular homogeneous and isotropic bounce. This is due to the fact that, even if the Hubble parameter itself is zero at the bounce point, the time derivative of the Hubble parameter is greater than zero. Hence, the equation of state (EoS) parameter,  $\omega$, of the universe is $\omega < -1$. This means that the total energy density disappears when the pressure is negative. From the phenomenological point of view, a stable bounce can be realized by considering a new physical mechanism in the matter sector, such as a quintom field \cite{Cai:2007qw}, a ghost condensate \cite{Lin2011}, or by changing the gravity theory at high-energy scales, even though it is difficult that a non-singular bounce occurs without the existence of ghost field. 
 
Cosmological bouncing scenarios including the ekpyrosis bounce and super-bounce have been studied \cite{Cai:2009rd, Cai:2014jla, Cai:2014bea}. Bouncing cosmological solutions have been considered in $ \mathcal{F}(R,T) $ gravity \cite{Shabani:2017kis}, and in $ \mathcal{F}(R) $, $ \mathcal{F}(T) $, $ \mathcal{F}(G) $ with $G$ the Gauss-Bonnet invariant and other types of modified gravities and also examined the dynamical stability of the solutions \cite{Bamba:2014zoa, Bamba:2016gbu, Odintsov:2015uca, Odintsov:2015zza, Bamba:2014mya}. 
It has been found that in modified gravity theories, the dynamics in the bouncing cosmology is substantially affected by the coupling parameter and the bouncing scale factors near the bouncing point \cite{Tripathy:2019nlw}. 
A summary of plausible explanation for the rapid expansion phase \cite{ Li:2009bn} has also been examined. Moreover, the bounce cosmology with an appearance of a Type-IV finite-time future singularity has been discussed in a ghost free Gauss-Bonnet gravity \cite{Odintsov:2022unp}. In addition, more kinds of theoretical aspects in bouncing cosmology have been studied \cite{Nojiri:2022xdoe, Odintsov:2020zct, Elizalde:2020zcb, Banerjee:2022gpy, Agrawal:2021msm, Agrawal:2022ppe}.

In this work, we explore the bouncing behavior of the universe in modified gravity with higher-order curvature terms in detail. It is shown that the extremal acceleration occurs at the bouncing point. It is also demonstrated that around the bouncing point, the universe universe behaves as it is filled with a perfect fluid. The organization of the present paper is as follows. In Sec. II, we briefly review the theoretical background of $ \mathcal{F}(R,T) $ gravity with higher-order curvature terms. Such a higher-order curvature term is a combination of a quadratic $R$-dependent term and a linear $T$-dependent term and it yields highly non-linear field equations. In Sec. III, we derive the solution of the gravitational field equations and analyze the bouncing criteria by applying the parametrization technique to the Hubble parameter. Finally, we summarize the results found in this work in Sec. IV.

\section{ Formulation of $ \mathcal{F}(R,T) $ gravity with higher-order curvature terms}
\qquad The higher order $ \mathcal{F}(R,T) $ gravity may be reviewed as a modification in basic gravitational theory that describes the evolution of the cosmos. The split-up model  $\mathcal{F}(R,T)=\mathcal{F}(R)+\mathcal{F}(T)$, where $\mathcal{F}(R)$ and $ \mathcal{F}(T)$ are arbitrary functions of $R$ and $T$, respectively, has gained significant attention as it allows researchers to investigate the contributions of each factor independently. The above split-up model has been employed by several authors \cite{Houndjo:2011tu,Mirza:2014nfa,Zaregonbadi:2016xna} to investigate cosmic dynamics from many perspectives.

The general action for $ \mathcal{F}(R,T)=\mathcal{F}(R)+\mathcal{F}(T) $ gravity \cite{Harko:2011kv} paired with the action of a matter field with matter Lagrangian $ \mathcal{S}_m $ is given by

\begin{equation} \label{1}
\mathcal{S}= \frac{c^4}{16\pi G}\int \mathcal{F}(R,T) \sqrt{-g}dx^4+\int \mathcal{S}_m\sqrt{-g} dx^4,
\end{equation}
where $R$ is the Ricci scalar and $T$ is the stress energy-momentum tensor trace. The following equation results from varying the action in Eq. (\ref{1}) concerning $g_{ij}$.
\begin{equation}\label{2}
    \mathcal{F}^{R}(R,T) R_{ij}-\frac{1}{2} g_{ij} \mathcal{F}(R,T)+(g_{ij}\Box-\nabla_i \nabla_j)\mathcal{F}^{R}(R,T)=8\pi G\, T_{ij}-\mathcal{F}^{T}(R,T)(T_{ij}+\Theta_{ij}),
\end{equation}
where the derivative of $\mathcal{F}(R,T)$ \textit{w.r.t.} $R$ and $T$ is represented by $\mathcal{F}^{R}(R,T)$ and $\mathcal{F}^{T}(R,T)$ respectively.
The d'Alembert operator $\Box$ is defined by $\Box= g^{ij}\nabla_{i} \nabla_{j}$, where $\nabla_i$ denotes the covariant derivative w.r.t. $g_{ij}$ related to the symmetric Levi-Civita connection.
The form of $\Theta_{ij}$ is as follows
\begin{equation}\label{3}
    \Theta_{ij}\equiv g^{lm} \frac{\delta T_{lm}}{\delta g^{ij}}= -2T_{ij}+g_{ij}S_m-2 g^{lm} \frac{\delta^2 S_m}{\delta g_{ij} \delta g^{lm}}.
\end{equation}
In this analysis we consider a perfect fluid in thermodynamic equilibrium and simply fix the matter Lagrangian $ \mathcal{S}_m=-p $ and consider the energy-momentum tensor of matter as
\begin{equation}\label{4}
T_{ij}=(p+\rho)u_i u_j-pg_{ij},
\end{equation}
where $ p $ and $ \rho $ denote the fluid's isotropic pressure and energy density, respectively. In a co-moving coordinate system, $ u^i=(0,0,0,1) $ is the four velocity vector which satisfies $ u^i u_i=1 $ and $ u^i\nabla_j u_i=0 $. Regarding (\ref{3}), the expression for the variation of the EMT of a perfect fluid is
\begin{equation}\label{5}
    \Theta_{ij}=-2T_{ij}-p g_{ij}.
\end{equation}
The gravitational equation of motion can be obtained by substituting Eq. (\ref{5}) into Eq. (\ref{2}) as
\begin{equation}\label{6}
\mathcal{F}^{R}(R,T) R_{ij}-\frac{1}{2}g_{ij} \mathcal{F}(R,T)+(g_{ij}\Box-\nabla_i \nabla_j) \mathcal{F}^{R}(R,T)= 8\pi G T_{ij}+\mathcal{F}^{T}(R,T)(T_{ij}+p g_{ij}).
\end{equation}
A more common representation of Eq. (\ref{6}) is as follows
\begin{eqnarray} \label{7}
G_{ij}=R_{ij}-\frac{1}{2}R g_{ij}
&=& \frac{8\pi G}{\mathcal{F}^{R}(R,T)}(T_{ij}+T_{ij}^{'}) ,
\end{eqnarray}
where 
\begin{equation}\label{8}
 T_{ij}^{'}=\frac{1}{8\pi G}\Big(\frac{1}{2} g_{ij}(\mathcal{F}(R,T)-R \mathcal{F}^{R}(R,T))+\Diamond_{ij} \mathcal{F}^{R}(R,T)+(T_{ij}+pg_{ij})\mathcal{F}^{T}(R,T)\Big) 
\end{equation}
and the operator $ \Diamond_{ij} $ is defined as $ \left(\nabla_i\nabla_j-g_{ij}\square \right)$.
We consider $ \mathcal{F}(R,T)= R+\lambda_1 R^2+2\lambda_2 T $, where $ \lambda_1 $ and $ \lambda_2 $ are constants. The function $ \mathcal{F}(R)=R+\lambda_1 R^2 $ was proposed by Starobinsky \cite{Starobinsky:1980te} as the first inflationary model, and function $ \mathcal{F}(T)=2\lambda_2 T $ is defined as a linear function of $ T $.

We consider spatially homogeneous and isotropic flat FLRW line element as
\begin{equation}\label{9}
ds^{2}=dt^{2}-a^{2}(t)\delta_{ij}dx^{i}dx^{j},   
\end{equation}
where $ a(t) $ is the scale factor. 

Taking $G=1$, the trace of the stress-energy tensor (\ref{4}) and scalar curvature are given by
\begin{equation}\label{10}
T=\rho -3p, \quad  R=-6(2H^2+\dot{H})
\end{equation}
where $ H = \dot{a}(t)/{a}(t) $ and the overhead dot indicates derivative \textit{w.r.t.} time $ t $. Using Eqs. (\ref{4}), (\ref{7}), (\ref{9}) and considering $ \mathcal{F}(R,T)=R+\lambda_1 R^2 +2\lambda_2 T $, we get the following field equations

\begin{equation}\label{11}
\left(8\pi+3\lambda_2\right)\rho-\lambda_2 p=3H^{2}+18\lambda_1\left(\dot{H}^{2}-6H^{2}\dot{H}-2H\ddot{H}\right),
\end{equation}
\begin{equation}\label{12}
\left(8\pi+3\lambda_2\right)p-\lambda_2 \rho = -2\dot{H}-3H^2+6\lambda_1(26\dot{H}H^2+2\dddot{H}+14H\ddot{H}+9\dot{H^2}).
\end{equation}
Thus from Eqs. (\ref{11}) and (\ref{12}), $p$ and $\rho$ can be written as
\begin{equation}\label{13}
p=\frac{(8\pi+3 \lambda_2)\rho-3H^2-18\lambda_1\left(\dot{H}^{2}-6H^{2}\dot{H}-2H\ddot{H}\right)}{\lambda_2},
\end{equation}
 
\begin{align}\label{14}
&\rho=\frac{\lambda_2 (-2 \dot{H}-3H^2+6\lambda_1(26 \dot{H}H^2+2 \dddot{H}+14H\ddot{H}+9\dot{H}^2 ))+(8\pi+3\lambda_2)(3H^{2}+18\lambda_1(\dot{H}^{2}-6H^{2}\dot{H}-2H\ddot{H}))}{\left(8\pi+2 \lambda_2 \right) \left(8\pi+4 \lambda_2 \right)}.
\end{align}
It can be observed that the pressure $p$ and energy density $\rho$ are dependent on the Hubble parameter $H$ and its time derivative $\dot{H}$.
In the current scenario, an additional constraint equation is needed to fully solve the system of field equations. This constraint equation is typically the state equation of a fluid. However, since we are specifically interested in a bouncing scenario, we impose a constraint on the Hubble parameter. Alternatively, one can also use a parametrization of physical or geometrical parameters to derive a constraint equation that is consistent with the system. This approach is known as a model-independent way to study DE models without compromising the underlying theory.

\section{ Bouncing solution and its dynamical behaviour}
The concept of the Big Bounce scenario is a kind of the possible way to avoid the issue of Big Bang singularity. In the Big Bounce scenario, the Big Bang appears as the beginning of a period of expansion that followed a period of contraction. In the scenario of a cyclic universe, the universe emerges from a previous contracting phase and eventually go through an expanding phase without any singularity. In other words, the universe undergoes a bouncing phase. The cosmological bounce can be regarded as an oscillatory or cyclic universe, in which one cosmological event is the result of the collapse of a previous phase. The following points are discussed in detail in order to construct a realistic bouncing model of the universe \cite{Ijjas:2016tpn,Cai:2007qw, Singh:2022gln}: 

\begin{enumerate}
\item In the contracting universe, the scale factor $ a(t) $ decreases ($ \dot{a}(t)< 0 $) and in the expanding universe, the scale factor increases ($ \dot{a}(t)>0 $). The derivative $ \dot{a}(t) = 0 $ or $ \ddot{a}(t)>0 $ in the vicinity of the bouncing point. When $ a(t) $ approaches a non-zero minimum value in the vicinity of the bouncing point, the model is called as a non-singular bouncing model.
\item Correspondingly, the universe contracts when $ H < 0 $, the universe expands when $ H > 0 $ and finally $ H=0 $ at
the time when the cosmological bounce occurs (which we call as the bouncing point). The condition $ \dot{H}= -4 \pi G \rho (1 + \omega)> 0 $, which is comparable to the violation of null energy condition ($ \rho+p>0 $), must be satisfied in the vicinity of the bouncing point for a successful bouncing model in standard cosmology. It can be seen from this equation that $ \omega \simeq -1 $ around the bouncing point. 
\item Another crucial aspect is that the equation of state (EoS) parameter $ \omega $ crosses the phantom divide line ($ \omega=-1 $) in the quintom model close to the bouncing point. 
\end{enumerate}
Given the aforementioned general criteria for a cosmological bounce and considering the Friedmann equations (\ref{11}) and (\ref{12}), it can be deduced that such a behavior can be achieved in the framework of $\mathcal{F}(R,T)$ cosmology. Specifically, by starting with a desirable form of the Hubble parameter $H(t)$, the corresponding scale factor $a(t)$ can be easily determined. In this paper, we focus on the cosmographic parameter $H(t)$, which describes the expansion of the universe and helps us find impressive bouncing solutions to the Einstein field equations.

\subsection{Technique of the reconstruction}
We analyze the Hubble parameter to examine the other cosmological quantities and investigate the cosmological behavior of the universe. Without loss of generality, the Hubble parameter $ H=\dot{a}/a$ is parametrized as
\begin{equation}\label{15}
H(t)=\alpha+ \beta t+ \gamma t^n,
\end{equation}
where $\alpha$, $\beta$, $\gamma$ and $ n $ are arbitrary constants. 
The motivation behind such a parametrization is to generate a bouncing scenario which gives the acceleration era of the universe together with the high rate of contraction before the bounce followed by the high rate of expansion after bounce by taking the suitable values of the model parameters.  

In Fig. 1, the bouncing position is shown in terms of the cosmic time $ t $ for the parameters $ \beta $, $ \gamma $, $ n=2.15 $ and $ \alpha =-10 $. The Hubble parameter evolves from its negative value to the positive one. This means that the universe evolves from a contraction  phase to the expansion one. The plot of $ H \sim t $ depicts a delay in the future bounce if $ \beta $ and $ \gamma $ decreases for fixed value of $ \alpha$ and $n$, respectively. The bouncing points occur at $ t \approx 3.41, 3.67$, and $3.93$ when $ \gamma=1.5 $ is fixed and $ \beta= -3.3, -4$, and $-4.7 $, respectively, as shown in Fig. 1(a). For $ \beta= -4.7 $, the bouncing points occur at $ t \approx 3.93, 3.25$, and $2.8 $ when $ \gamma=  1.5, 2$, and $ 2.5 $, respectively, as shown in Fig. 1(b). We consider the bouncing point $ t \approx 3.93 $ when $ \beta=-4.7 $ and $ \gamma=1.5 $. The nature of Hubble parameter is shown in Table I for the different values of model parameters $ \beta $ and $ \gamma $ and fixed values of $ n $ and $ \alpha $. Here, the standard unit of $H$ is taken as $km/s/Mpc$ and $t$ as $Gyr$. 

By using Eq. (\ref{15}), the scale factor $ a(t) $ can be expressed as 
\begin{equation}\label{16}
a(t)=k e^{\frac{\gamma  t^{n+1}}{n+1}+\frac{\beta  t^2}{2}+\alpha  t}.
\end{equation}
where $k$ is an integration constant. 

In Fig. 2(b), the evolution of the scale factor is shown. At the instant of the bounce, $t=3.93$, the scale factor behaves symmetrically. 
As a result, the universe evolves from a former contraction phase $(t < 3.93)$ to a later expansion one $(t > 3.93)$. 
At the bouncing point $t = 3.93$, the scale factor does not vanish. 
For a successful bounce, as demonstrated in Fig. 2, 
the Hubble parameter have to be negative before the bounce and positive after the bounce with $ \dot{H}= -4 \pi G \rho (1 + \omega)> 0 $ close to the bouncing epoch. 

\begin{figure}\centering
	\subfloat[]{\label{a}\includegraphics[scale=0.41]{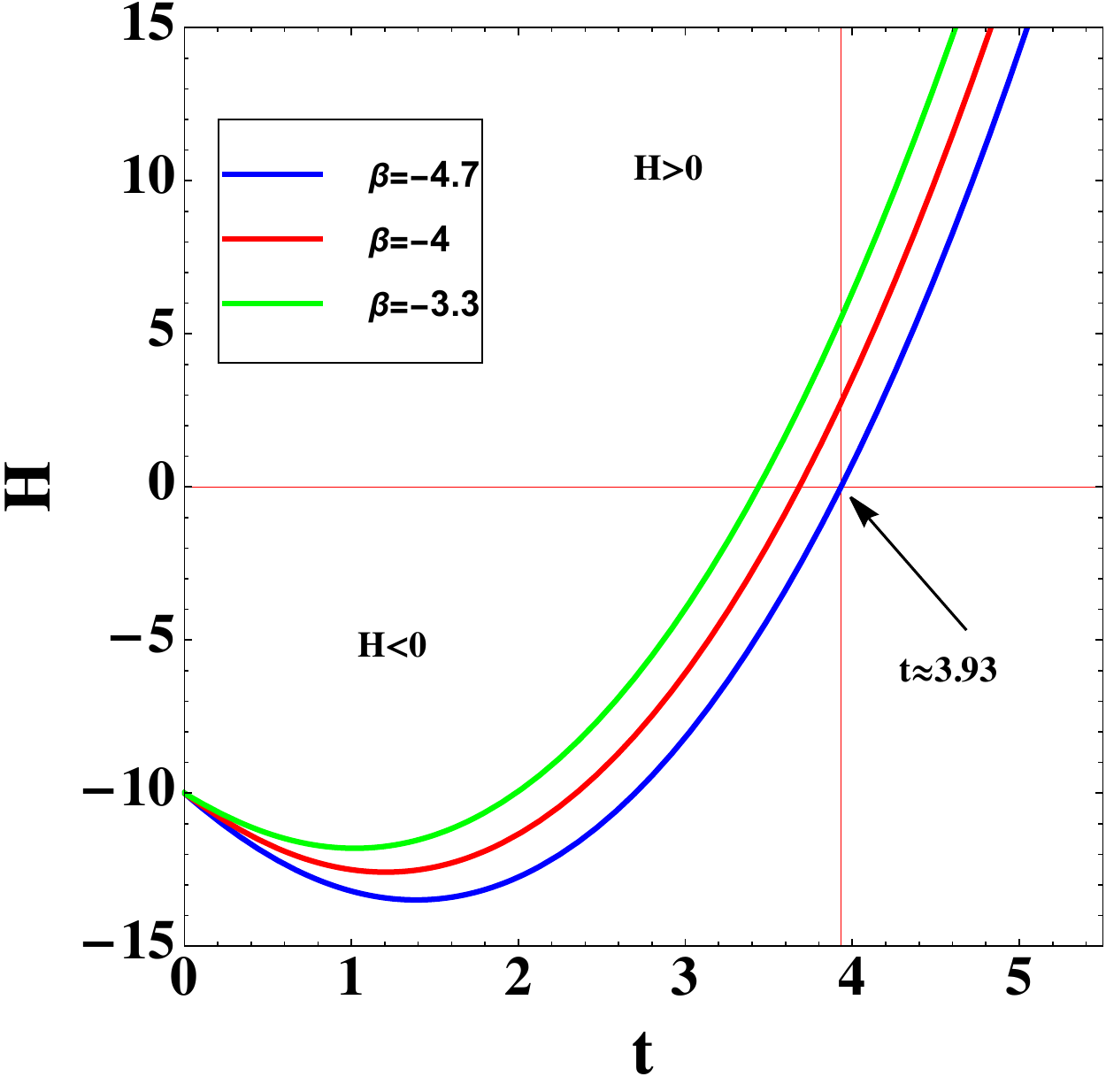}}\hfill
	\subfloat[]{\label{b}\includegraphics[scale=0.40]{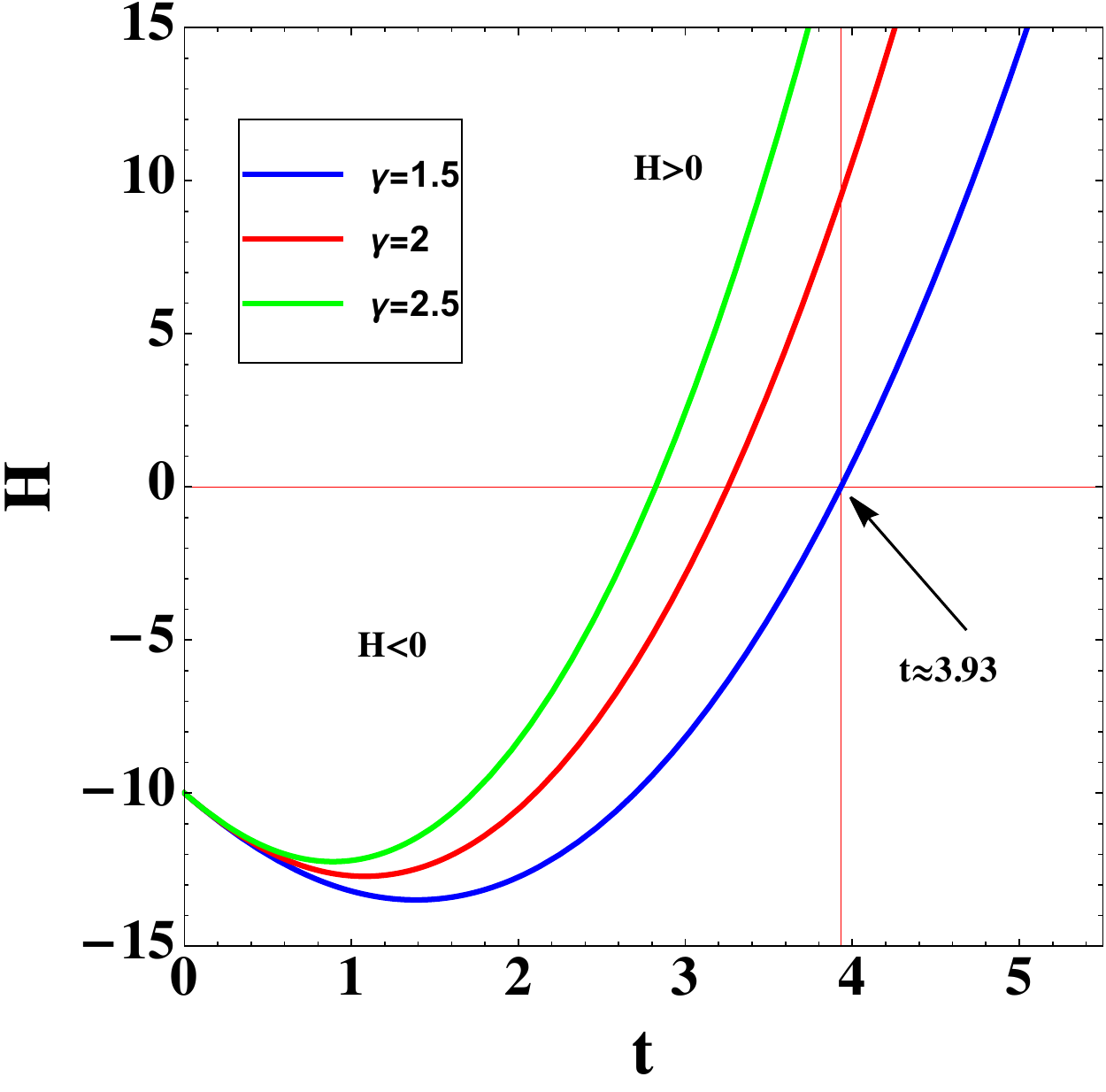}} 
	\caption{ Evolution of the Hubble parameter $ H $ as a function of the cosmic time $t$ for distinct values of $ \beta $ and $ \gamma $  respectively.}
\end{figure}

\begin{table}[htbp]
\caption{ \textbf{Constraints on the model parameters}}
\begin{center}
\label{tabparm}
\begin{tabular}{ | p{2cm} | p{2cm} | p{3cm} | p{4cm} | }
 \hline
 \multicolumn{4}{|c|}{\textbf{ Fig. 1(a) }}
 \\
 \hline
 $ ~~\beta $ &  ~~ $ \gamma $ & ~~  Time (Gyrs) & Nature of H(km/s/Mpc)
 \\
 \hline

  ~~$ -4.7 $ & ~~$ 1.5 $ & ~~$  0< t < 3.93 $ & ~~  \textit{Contracting } 
\\
 ~~$ -4 $ & ~~$ 1.5 $ & ~~$  0< t < 3.67 $ & ~~   \textit{Contracting } 
\\
  ~~$ -3.3 $ & ~~$ 1.5 $  & ~~$  0< t < 3.41 $ & ~~   \textit{Contracting }   
\\
\hline
  ~~$ -4.7 $ & ~~$ 1.5 $ & ~~$   3.93< t < \infty $ &  ~~  \textit{Expanding}
\\
  ~~$ -4 $ & ~~$ 1.5 $ & ~~$   3.67 < t < \infty $ &  ~~  \textit{Expanding}
\\
~~$ -3.3 $ & ~~$ 1.5 $  & ~~$   3.41< t < \infty $ &  ~~ 
 \textit{Expanding}
\\ 
 \hline

 \multicolumn{4}{|c|}{\textbf{\textbf{ Fig. 1(b) }}} 
 \\
 \hline
  $ ~~\beta $ &  ~~ $ \gamma $ & ~~ Time (Gyrs) & Nature of H(km/s/Mpc)
 \\
 \hline

 ~~$ -4.7 $ & ~~$ 1.5 $ & ~~$ 0< t < 2.8 $ & ~~  \textit{Contracting }
\\
  ~~$ -4.7 $ & ~~$ 2 $ & ~~$ 0< t < 3.25 $ & ~~  \textit{Contracting }  
\\
~~$ -4.7 $ & ~~$ 2.5 $  & ~~$ 0< t < 3.93 $ &  ~~  \textit{Contracting }  
\\
\hline
~~$ -4.7 $ & ~~$ 1.5 $  & ~~$ 2.8< t < \infty $ &  ~~ \textit{ Expanding}
\\
 ~~$ -4.7 $ & ~~$ 2 $  & ~~$ 3.25 < t < \infty $ &  ~~  \textit{ Expanding} 
\\
  ~~$ -4.7 $ & ~~$ 2.5 $  & ~~$ 3.93 < t < \infty $ & ~~  \textit{ Expanding}   
\\ 
 \hline
\end{tabular}
\end{center}
\end{table}

The deceleration parameter $ q $ is defined as 
\begin{equation}\label{17}
q=-\frac{a\ddot{a}}{\dot{a}^2}=-1-\frac{\dot{H}}{H^2},
\end{equation}
By using Eqs. (\ref{15}), (\ref{17}), the expression for the deceleration parameter $q$ is expressed as
\begin{equation}\label{18}
q=-1 - \frac{\beta + n \gamma t^{n-1} }{(\alpha + \beta t + \gamma t^n)^2}.
\end{equation}
\begin{figure}\centering
	\subfloat[]{\label{a}\includegraphics[scale=0.42]{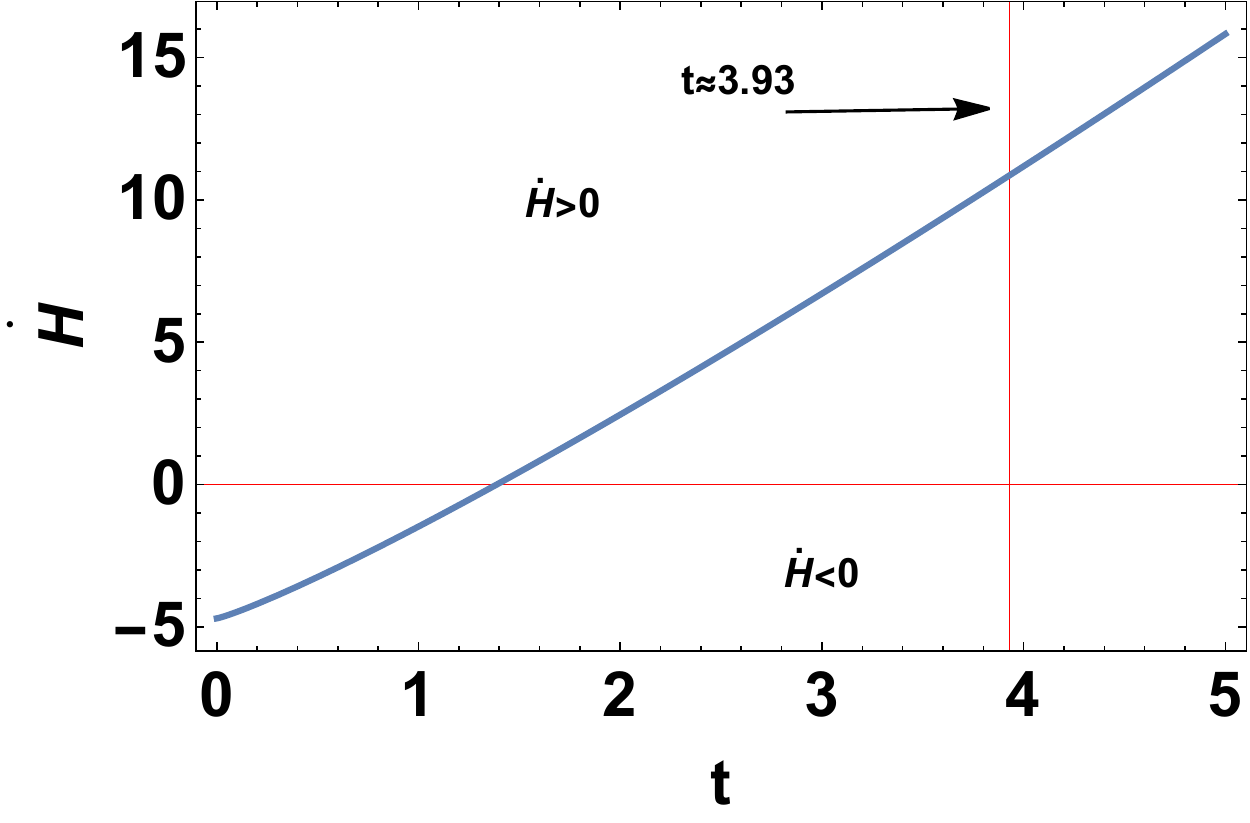}}\hfill
	\subfloat[]{\label{b}\includegraphics[scale=0.42]{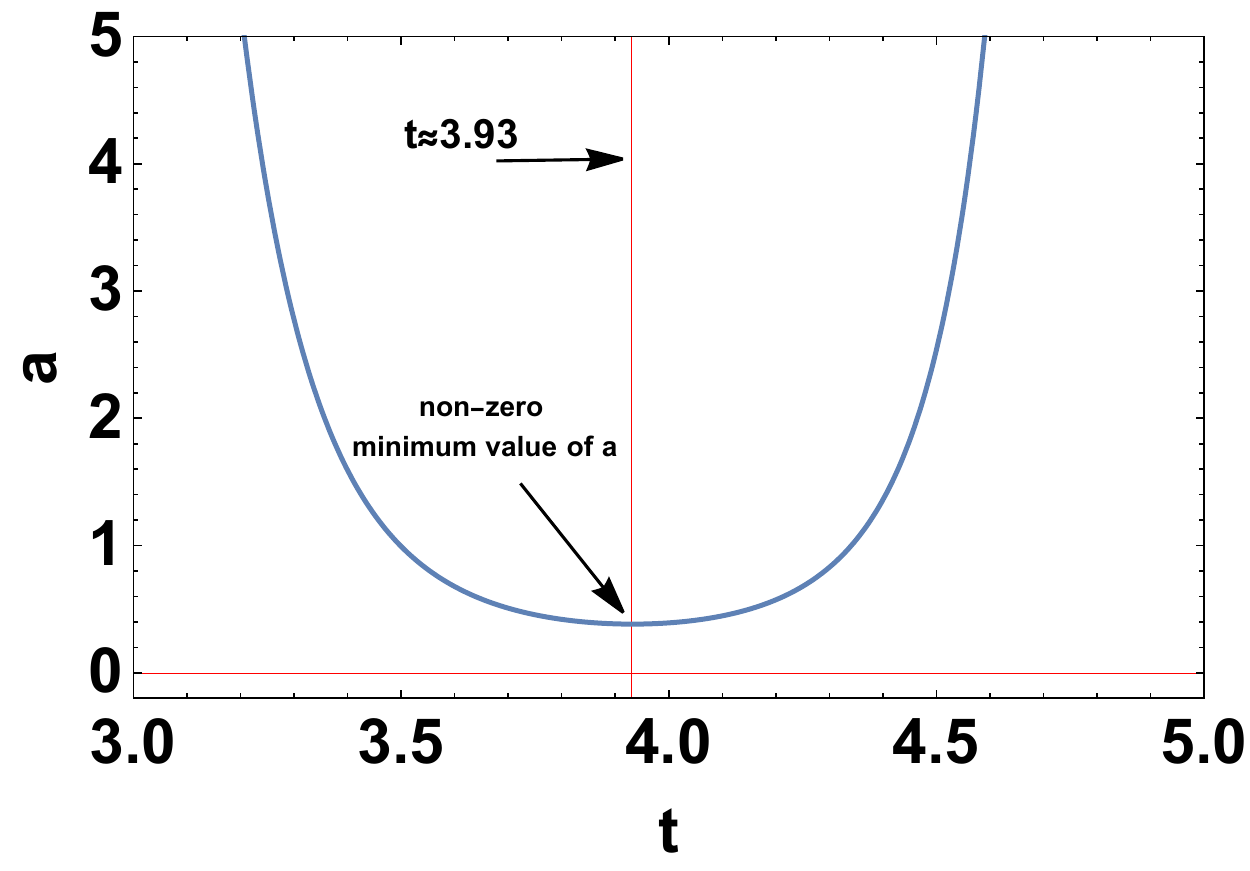}}
	\caption{ Evolution of the time derivative of the Hubble parameter $ H $ and that of the scale factor $a$ as a function of the cosmic time $t$.}  
	
\end{figure} 

\begin{figure}\centering
   \includegraphics[scale=0.5]{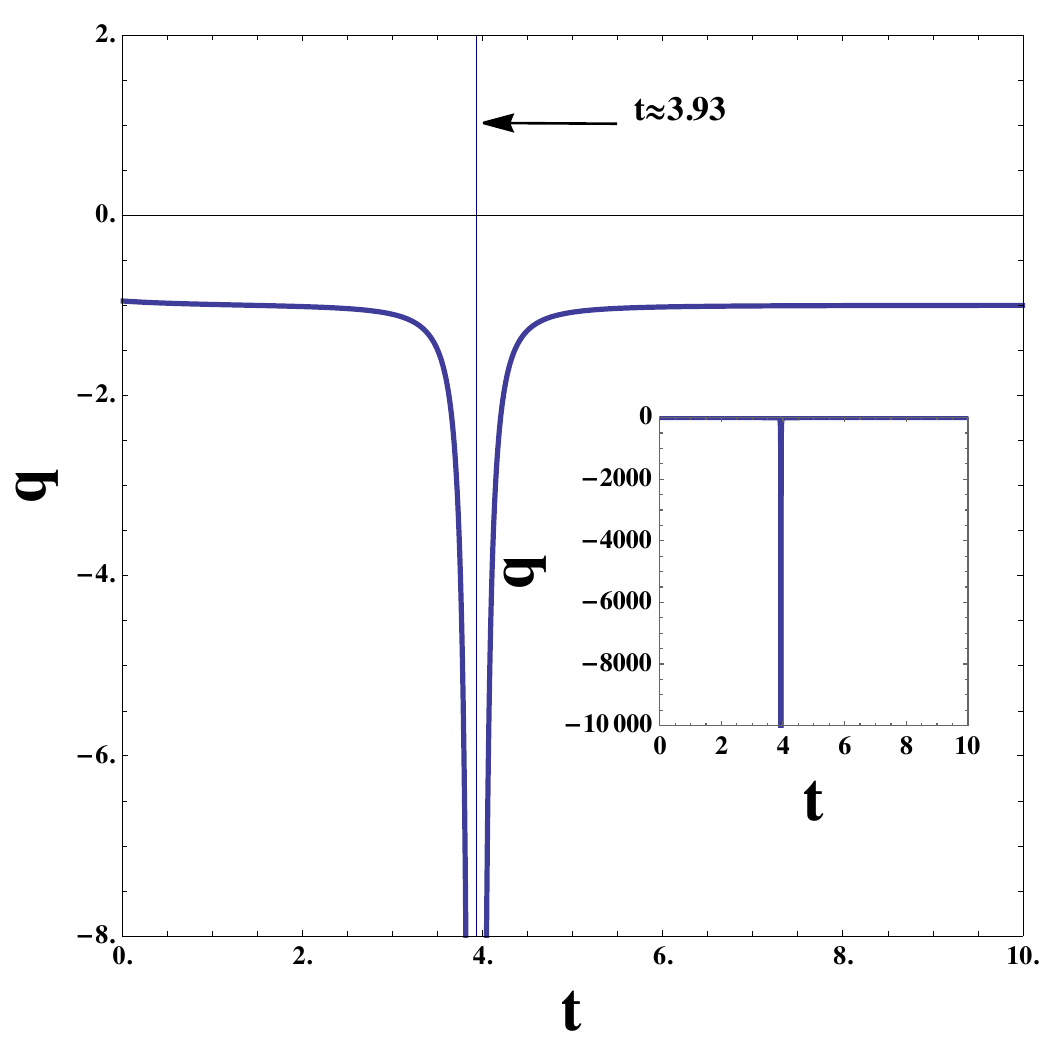} 
	\caption{ Evolution of the deceleration parameter $ q $ as a function of the cosmic time $t$.}
\end{figure}

The expansion of the model depends on the deceleration parameter $ q $. A positive value of $ q $ represents a decelerated universe whereas a negative value of $ q $ explains an accelerating model of the universe. Fig. 3 depicts the symmetrical nature of the deceleration parameter in the neighbourhood of the bouncing point $ t=3.93 $. It is significant to notice that before and after the bouncing point the deceleration parameter has a constant value of $ -1 $ and has a negative value for both contracting and expanding universes. From Eq. (\ref{17}), it has been seen that the deceleration parameter $ q<0 $, $ \forall t $ and the acceleration is extreme at the bouncing point $ t \approx 3.93 $ (see Fig. 3). The authors of  \cite{Singh:2017ubs} investigated the many forms of growth that our universe shows including i) $ q < -1 $; super exponential expansion, (ii) $ -1 \leq q < 0 $; exponential expansion ( $ q= -1 $ also known as de-Sitter expansion), (iii) $ q=0 $; expansion with constant rate, (iv) $ -1 < q < 1 $; accelerating power expansion, and (v) $ q>0 $ ; decelerating expansion. In other words, one can say that this bouncing model is an ever-accelerating model. Table II exhibits the nature of $ q $ in the evolution of the Universe for fixed values of free model parameters.

\begin{table}[htbp]
\caption{ \textbf{Evolution of the universe for $ n=2.15 $, $\alpha=-10$, $ \beta=-4.7 $ and $ \gamma=1.5 $.}}
\begin{center}
\label{tabparm}
\begin{tabular}{l c c c r} 
\hline\hline
\\ 
\quad & $ q $~~~   & ~~~$ H $~~~ & ~~ \textit{Time interval} & \qquad \textit{Evolution of Universe}
\\
\hline 
\\
\quad & $ <0 $ & $ <0 $ &  ~~$ 0< t < 3.93 $  &  $ q \oplus H $\footnote{ Accelerated contraction}
\\
\quad & $ <0 $ & $ =0 $ & $ ~~t \approx 3.93 $ &   $ q_{max}\oplus H $\footnote{ Bounce with max. acceleration} 
\\
\quad & $ <0 $ & $ >0 $ & ~~$ 3.93< t < \infty $ & $ q \oplus H $\footnote{ Accelerated Expansion}
\\
\hline 
\\
\end{tabular}    
\end{center}
\end{table}

We solve the effective field equations explicitly and define the EoS parameter as $ \omega=p/\rho $. The energy density $ \rho $ and pressure $p$ of the universe are given by

 \begin{multline}\label{19}
\rho= \frac{1}{8 \left(\lambda _2+2 \pi \right) \left(\lambda _2+4 \pi \right)}\Bigg[[-2(\beta+nt^{n-1}\gamma)-3(\alpha+\beta t+\gamma t^n)^2+6[2(n-2)(n-1)nt^{n-3}\gamma+9(\beta+nt^{n-1}\gamma)^2+\\
14n(n-1)t^{n-2}\gamma(\alpha+\beta t+\gamma t^n)+26(\beta+nt^{n-1}\gamma)(\alpha+\beta t+\gamma t^n)^2]\lambda_1]\lambda_2-3[(\alpha+\beta t+\gamma t^n)^2
 +\\6[(\beta+nt^{n-1}\gamma)^2-2n(n-1)t^{n-2}\gamma(\alpha+\beta t+\gamma t^n)
 -6(\beta+nt^{n-1}\gamma)(\alpha+\beta t+\gamma t^n)^2]\lambda_1](8\pi+3\lambda_2)\Bigg],
\end{multline}
 and 
\begin{multline}\label{20}
p=\frac{1}{\lambda_2}\Bigg[-3(\alpha+\beta t+\gamma t^n)^2-18[(\beta+nt^{n-1})^2-2n(n-1)t^{n-2}\gamma(\alpha+\beta t+\gamma t^n)-6(\beta+nt^{n-1}\gamma)(\alpha+\beta t+\gamma t^n)^2]\lambda_2+ \\(8\pi+3\lambda_2)\rho\Bigg].
\end{multline}
 
 \begin{figure}\centering
	\subfloat[]{\label{a}\includegraphics[scale=0.38]{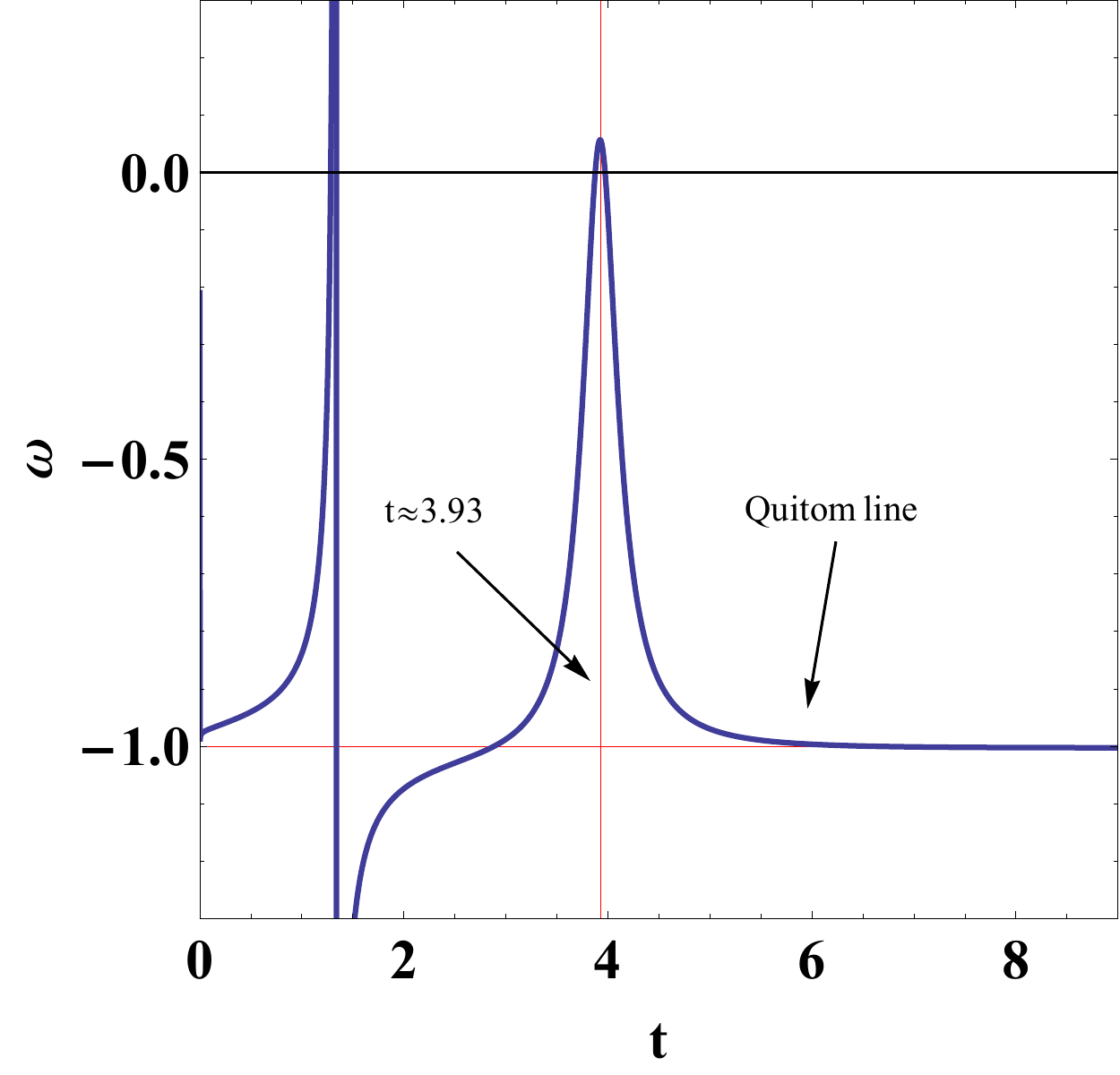}}\hfill
	\subfloat[]{\label{b}\includegraphics[scale=0.46]{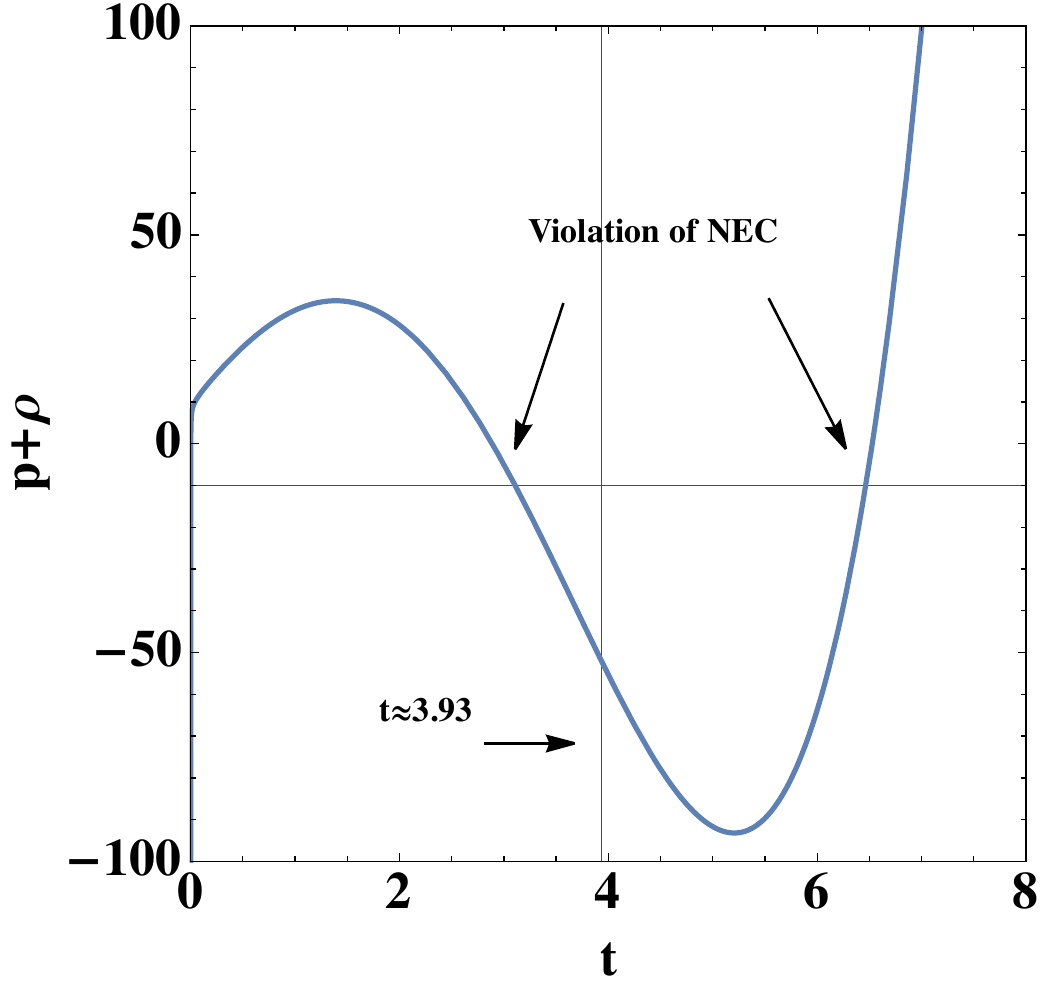}}
	\caption{ Evolution of the equation of state (EoS) parameter $ \omega $ and the Null energy condition $ (NEC: p+\rho) $ as a function of the cosmic time $t$.}  
	
\end{figure} 
 
We plot Fig. 4 for $ \beta=-4.7 $, $ \gamma=1.5 $ and the constants $ \lambda_1=-0.05$, $\lambda_2=-7.8 $ in the action of $\mathcal{F}(R,T)$ gravity. It can be seen from Fig. 4 that the EoS parameter $ \omega $ crosses the phantom divide line ($ \omega= -1 $) and shows the perfect fluid behavior in the short-range $ 2.95<t<5.62 $ in the vicinity of the bouncing point $ t \approx 3.93 $.
From Figs. 2, 3 and 4, we observe the following: 
\begin{enumerate}
\item In Fig. 2(a), $\dot{H}$ evolves from negative to positive and in the neighbourhood of the bouncing point $ t \simeq 3.93 $, $ \dot{H} > 0. $  

\item The scale factor $ a(t) $ decreases during the contraction of the universe ($ \dot{a}(t)<0 $) before the bounce and increases during the expansion of the universe ($ \dot{a}(t)>0 $) after the bounce at $ t \simeq 3.93$. Moreover, the universe starts from finite volume as $\lim_{t \to 3.93} a(t) \to 0.38264 $ at the bounce $ t \simeq 3.93 $. Thus, the bounce without a singularity can be realized as shown in Fig. 2(b). In addition, the fact that the value of the Hubble parameter vanishes at the transfer point is guaranteed by the point that the value of the scale factor is finite at $t = 3.93$.

\item Fig. 3 illustrates that the  model is an ever accelerating one and attains extremum value at the bouncing point $t = 3.93$. 

\item The value of the EoS parameter $ \omega $ evolves from $ \omega< -1 $ to $ \omega= -1 $ in the neighbourhood of the bouncing point $ t \simeq 3.93 $. $ \omega $ crosses the line of the phantom divide $ \omega= -1 $ as depicted in Fig. 4(a). In Fig. 4(b), the fact that the null energy condition ($\rho + p >0 $) is satisfied is shown. The null energy condition violates in the range $ 3.086 \leq t \leq 6.448 $, in which the bouncing conditions are met.

\end{enumerate}

\subsection{Violation of energy conditions}
In general relativity, the energy conditions (ECs) are crucial to investigate the issue of the singularity of the space-time and the behavior of the null, space-like, time-like or light-like geodesics. By exploring these conditions, 
the nature of cosmic geometries and its relations to the stress energy-momentum, which must be positive, can be found. These conditions are just simple constraints on various linear combinations of the energy density and pressure.  This results in the fact that energy density cannot be negative and that gravity always exhibits an attractive force.

 ECs impose restrictions on the ability of the stress tensor to contract at each location of the space \cite{Curiel:2014zba}. Conditions for various theories can also be written in a geometric form in addition to their original physical form that complies with the gravitational field equation by taking the Ricci tensor as the stress tensor.
  The four main ECs are the weak energy condition (WEC), the strong energy condition (SEC), the dominant energy condition (DEC) and the null energy condition (NEC).
 ECs in different forms are given in Table III, where $t^i$ and $\xi^j$ are co-oriented time-like vectors and $k^i$ is a null (light-like) vector. These conditions cannot be independent of each other \cite{Kontou:2020bta}. 

\begin{table}[htbp]
\caption{ \textbf{Energy Conditions}}
\begin{center}
\label{tabparm}
\begin{tabular}{ | p{2 cm} | p{3.5cm} | p{3cm} | p{6cm} | }
 \hline
 \multicolumn{4}{|c|}{\textbf{ Energy Conditions }}
 \\
 \hline
  ~~Condition~~   &   ~~Physical Form     &   ~~Geometric Form  &    ~~Perfect Fluid     
 \\
 \hline
  ~~ $ WEC $ ~~ & ~~ $ T_{ij}t^it^j\geq 0 $  & ~~$  G_{ij}t^it^j\geq 0 $  & ~~$  \rho \geq 0, \rho+p \geq 0 $  
\\
\hline
 ~~ $ SEC $ ~~ & ~~  $ (T_{ij}-\frac{T}{n-2}g_{ij})t^it^j\geq 0  $ & ~~ $ R_{ij}t^it^j\geq 0 $  & ~~  $ \rho+p \geq 0, (n-3)\rho+(n-1)p \geq 0 $
\\
\hline
  ~~ $ DEC $~~  & ~~ $ T_{ij}t^i\xi^j\geq 0 $  & ~~ $ G_{ij}t^i\xi^j\geq 0 $ & ~~   $\rho \geq |p| $  
\\
\hline
  ~~ $ NEC $ ~~ & ~~ $ T_{ij}k^ik^j\geq 0 $ & ~~$ R_{ij}k^ik^j\geq 0 $  &  ~~ $ \rho+p \geq 0 $
\\
\hline
\end{tabular}    
\end{center}
\end{table}

\begin{figure}\centering
	\subfloat[]{\label{a}\includegraphics[scale=0.44]{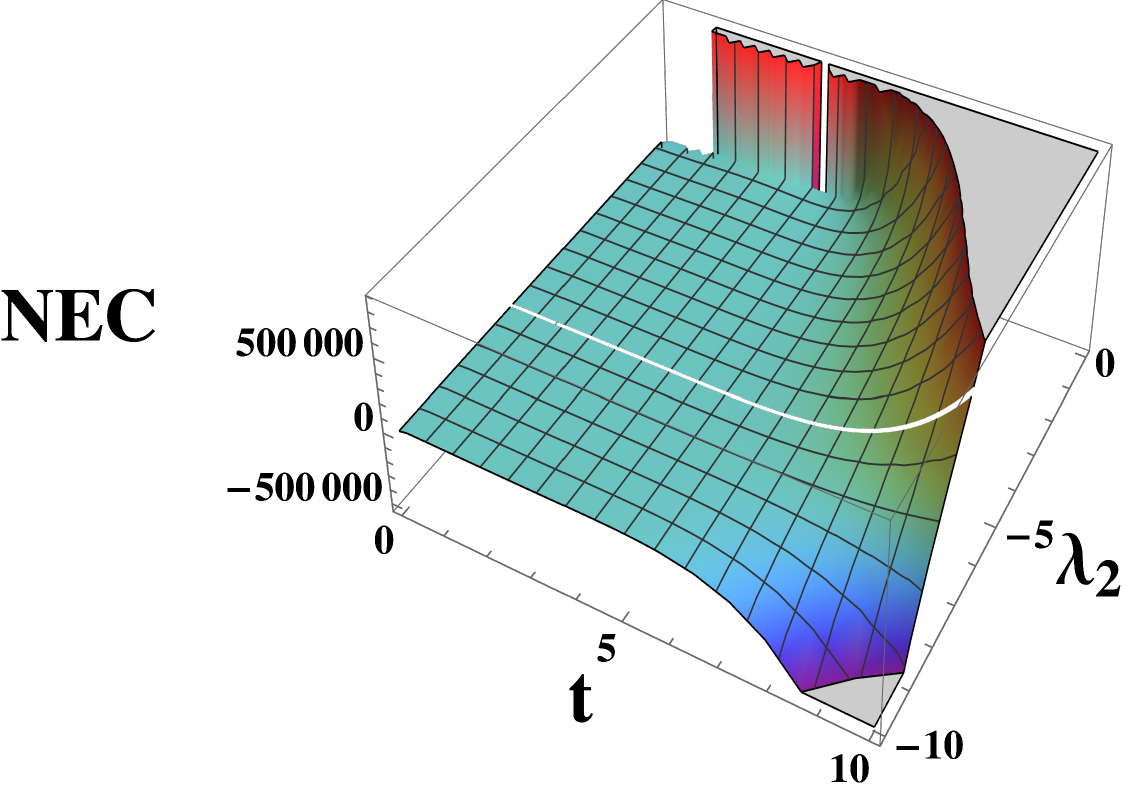}}\hfill
	\subfloat[]{\label{b}\includegraphics[scale=0.40]{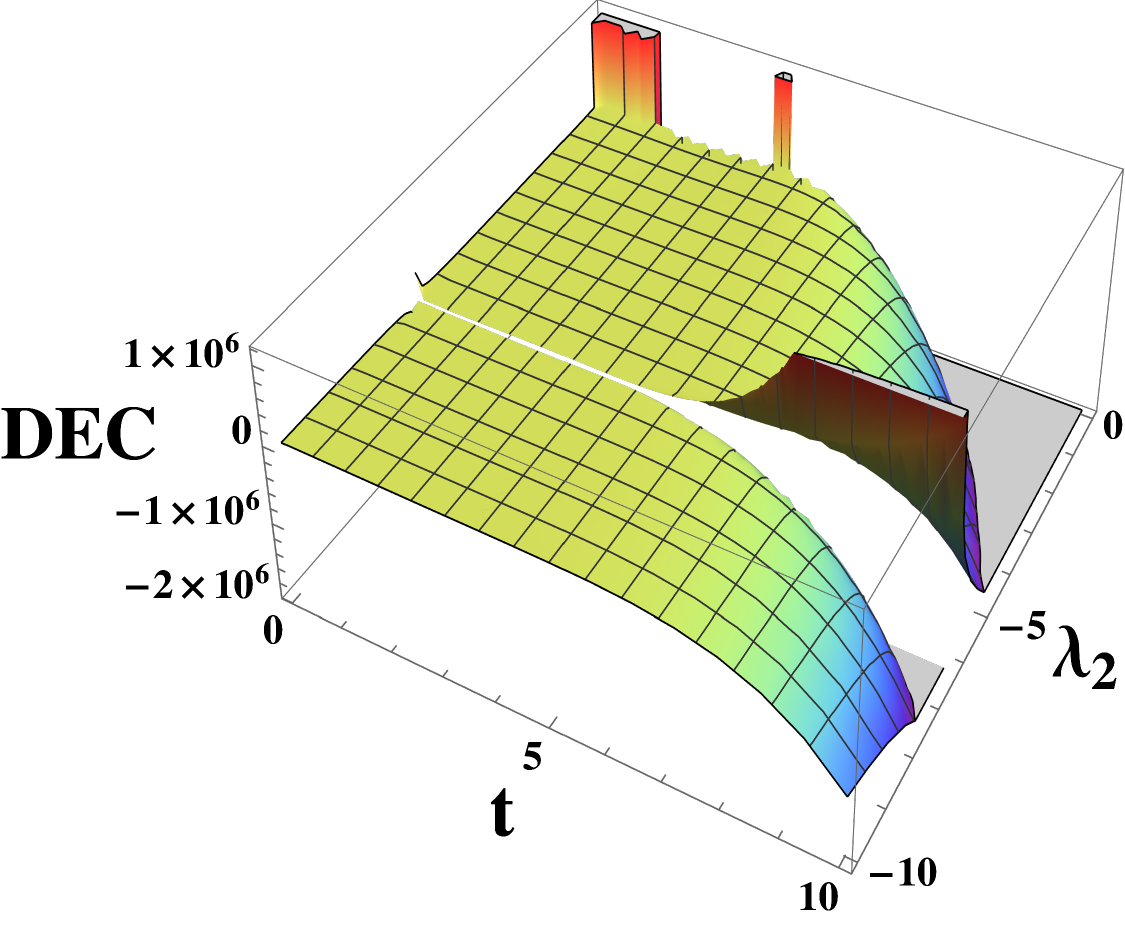}}\par 
	\subfloat[]{\label{c}\includegraphics[scale=0.44]{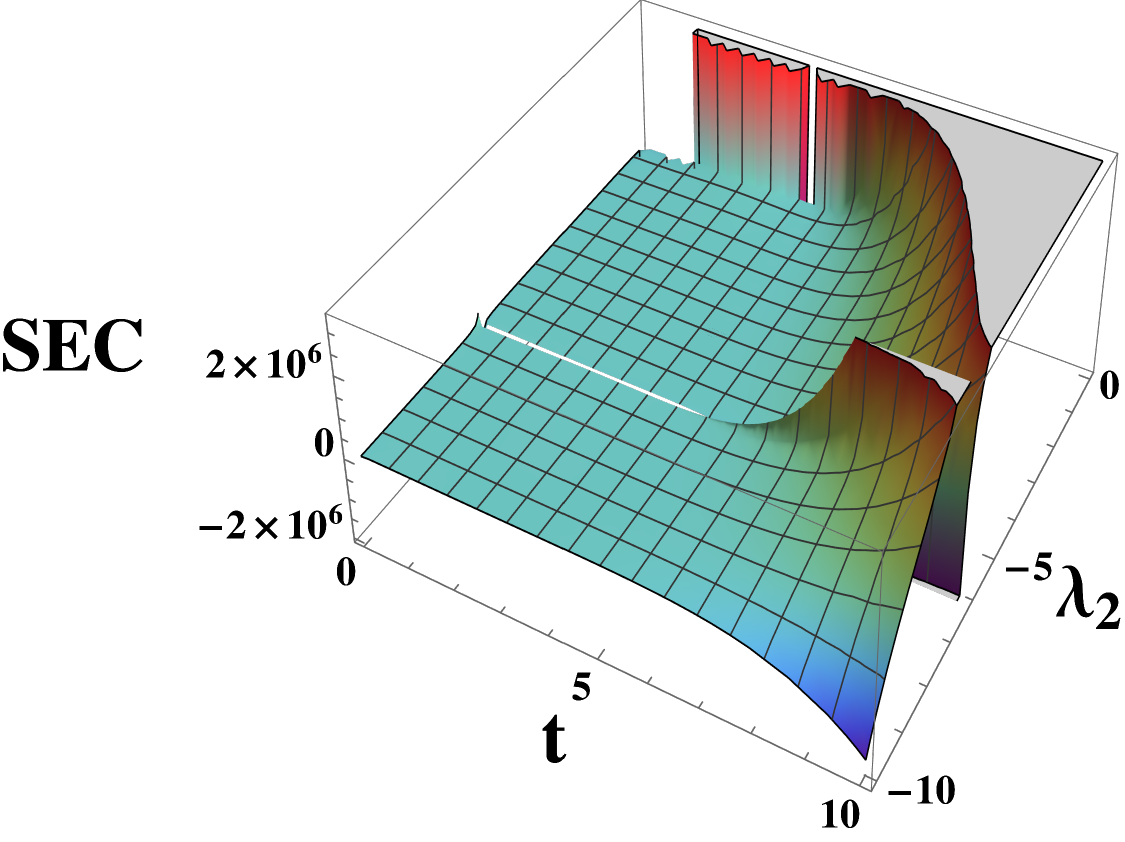}}
	\caption{\scriptsize Behaviors of NEC, DEC and SEC for $ n=2.15 $.}
\end{figure}
 
The aforementioned relationships hold as: WEC  $ \Rightarrow $ NEC; SEC   $\Rightarrow$ NEC;  DEC  $\Rightarrow$ NEC. If the NEC is violated, none of the ECs can be satisfied \cite{Visser:1995cc}. It is considered that the NEC is met for all of the stable systems. In a sensible system, other ECs can be violated by adding a suitable (positive or negative) cosmological constant. In the event that the stress-energy tensor violates the NEC, the system becomes catastrophically unstable, meaning that a ghost is present. This occurs when the kinetic term has the \textit{wrong} sign, or when the system contains \textit{tachyons} \cite{Caldwell:1999ew}, which are modes that expand exponentially with arbitrarily short wavelengths. Modified gravity theories deviating from general relativity can be viewed as having a background scalar field with a stress-energy tensor, serving as a sort of cosmological constant. This means that in such theories, it is possible for the NEC to be violated. Consequently, modified gravity theories may result in degenerate dispersion relations, enabling the breaking of the NEC.


The NEC plays a crucial role among various ECs considered in general relativity. In terms of the NEC, the energy-momentum tensor for matter $ T_{ij} $ satisfies $ T_{ij}k^i k^j $, $ \forall $ $ k^i $, \textit{i.e.}, for any vector with $ g_{ij}k^i k^j=0 $.
There are two main reasons why the NEC is considered significant \cite{Rubakov:2014jja}. First, it was previously believed that the NEC could only be violated in a theory involving a scalar field with non-minimal coupling to gravity \cite{Flanagan:1996gw}. Second, the NEC is a fundamental assumption in the Penrose singularity theorem \cite{Penrose:1964wq}, which is applicable in general relativity. The Penrose singularity theorem requires two conditions: (i) the NEC be upheld; (ii) the Cauchy hypersurface be non-compact. The theorem states that if there is a trapped surface in space, a singularity will eventually occur. A trapped surface is a closed surface where outgoing light rays are actually converging (moving inward).

%

In Fig. 5, it is demonstrated how the energy criteria are violated. It is shown that there is no singularity close to the bouncing epoch. The circumstances of energy change in the vicinity of the bouncing point $ t \simeq 3.93 $. According to the anticipated matter bounce scenario, the energy conditions $ \rho+p $ and $\rho+3p $ become negative near the bounce, and therefore there is a strong evidence of the violation of the energy conditions. The violation of the energy conditions causes the phantom phase with $ \omega\leq-1 $.

\subsection{Scalar field description}
 
A model with both the phase of quintessence ($ -1 < \omega <-1/3 $) and that of phantom ($ \omega <-1 $) is referred to as quintom model. 
The necessary condition of the present model to be consistent with observations is $ \dot\phi^2 << V(\phi) $. This implies that the kinetic energy (KE) of the scalar field is minimal in comparison with its potential energy (PE). There have been proposed a number of models with $ \omega\simeq-1 $ to explain the cosmic acceleration (including inflation in the early universe). \\

The action in general relativity is described as 
\begin{equation}\label{21}
S=\frac{c^4}{16\pi G}\int{R\sqrt{-g}d^4x}+S_m .
\end{equation}
where $ S_m $ consists of the action for the quintessence-like scalar field ($ S_{m_q} $) and that for phantom-like one $ (S_{m_{ph}})$, which are given by
\begin{equation}\label{22}
S_{m_q}=\int{\Big[-\frac{1}{2}\partial_\mu \phi_q \partial^\mu \phi_q -V(\phi_q)\Big]\sqrt{-g}d^4x},
\end{equation}
and
\begin{equation}\label{23}
S_{m_{ph}}=\int{\Big[\frac{1}{2}\partial_\mu \phi_{ph} \partial^\mu \phi_{ph} -V(\phi_{ph})\Big]\sqrt{-g}d^4x},
\end{equation}
respectively. 

The energy density $\rho_{\phi_q},\rho_{\phi_{ph}} $ and pressure $ p_{\phi_q},p_{\phi_{ph}} $ for the quintessence-like and phantom-like scalar fields are represented as \cite{Singh:2018xjv, Barrow:1988xh}
\begin{equation}\label{24}
\rho_{\phi_q}=\frac{1}{2}\dot{\phi_q}^2+V(\phi_q), \qquad p_{\phi_q}=\frac{1}{2}\dot{\phi_q}^2-V(\phi_q)
\end{equation}
\begin{equation}\label{25}
\rho_{\phi_{ph}}=-\frac{1}{2}\dot{\phi_{ph}}^2+V(\phi_{ph}), \qquad p_{\phi_{ph}}=-\frac{1}{2}\dot{\phi_{ph}}^2-V(\phi_{ph})
\end{equation}
The subscripts of $ q $ and $ ph $ denote the quintessence-like and phantom-like scalar fields, respectively. 
In Appendix, we express the kinetic energy and potential energy of the quintessence-like ($ \frac{1}{2}\dot\phi^2_q $) scalar field and that of the phantom-like ($ \frac{1}{2}\dot\phi^2_{ph} $) one. For the phantom-like and quintessence-like scalar fields, the EoS parameters $ \omega $ are represented as \cite{Singh:2018xjv}

\begin{equation}\label{29}
\omega=\frac{p_{\phi_{ph}}}{\rho_{\phi_{ph}}}<-1,\,\,\,\omega=\frac{p_{\phi_q}}{\rho_{\phi_q}}>-1.
\end{equation}

Eqs. (\ref{26}), (\ref{27}) and (\ref{28}), which are presented in Appendix, yield 
\begin{equation}\label{30}
\dot{\phi}_{q}^{2}= \dot{\phi}_{ph}^{2}. 
\end{equation}
Eq. (\ref{30}) indicates that the EoS parameter crosses the phantom divide line of $\omega = -1$. This is the significant criteria of the bouncing model and compatible with the consequences in Ref. \cite{Cai:2007qw}. Thus, it is interpreted that the present model is a non-singular bouncing model in $ \mathcal{F}(R,T) $ gravity. 

\begin{figure}\centering
	\subfloat[]{\label{a}\includegraphics[scale=0.72]{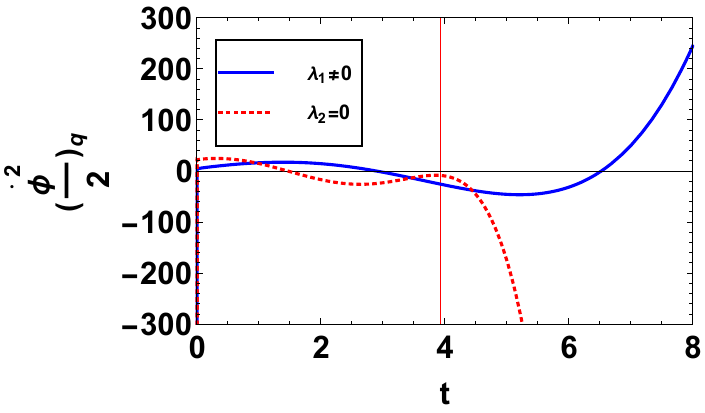}}\hfill
	\subfloat[]{\label{b}\includegraphics[scale=0.72]{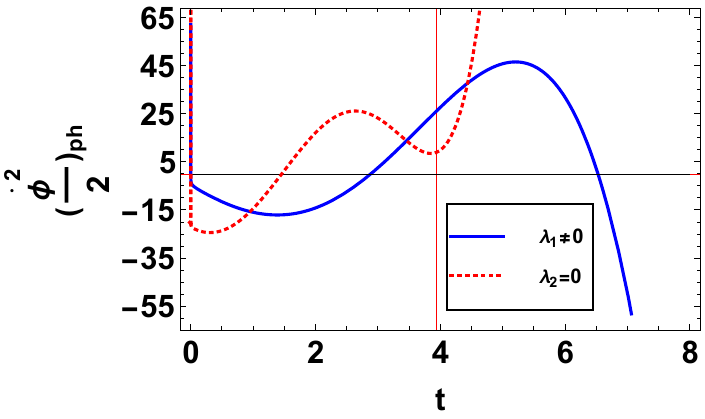}}\par 
	\subfloat[]{\label{c}\includegraphics[scale=0.75]{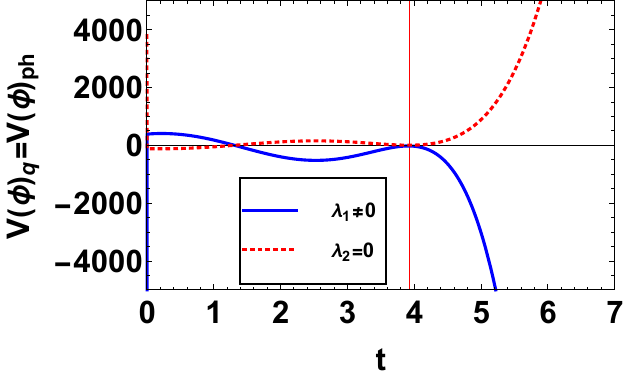}}
\caption{\scriptsize The evolution of the kinetic energy of the quintessence-like field $ (\frac{1}{2}\dot{\phi_q}^2 ) $ and the phantom-like field $ (\frac{1}{2}\dot{\phi_{ph}}^2) $ and that of the potential energy $ V(\phi) $ with respect to the cosmic time $ t $ for quintessence and phantom-like scalar fields in $ \mathcal{F}(R,T) $ gravity ($ \lambda_1\neq 0 $) and GR ($ \lambda_2=0 $).}
\end{figure}

A unified model for dark energy and dark matter can be constructed as a development of properties of a scalar field with its negative kinetic energy. From Eq. (\ref{25}), it is understood that the kinetic energy of the phantom-like scalar field is given by $ (\frac{1}{2}\dot{\phi_{ph}}^2)=-(\rho_{\phi_{ph}}+p_{\phi_{ph}}) $. Various cosmological models of scalar field with negative kinetic energy have been studied \cite{Gibbons:2003yj,Nojiri:2003ag,Caldwell:1999ew,Cline:2003gs} to account for the late-time cosmic acceleration. In Fig. 6, we show the inequality of the kinetic energy of the quintessence-like filed ($ \frac{1}{2} \dot{\phi}_{q}^{2} $) and that of the phantom-like ($ \frac{1}{2} \dot{\phi}_{ph}^{2} $) one and the potential energies $ V(\phi _q) $ and $ V(\phi_{ph}) $ of the scalar fields with respect to the cosmic time $ t $ in $ \mathcal{F}(R,T) $ gravity as well as general relativity.

The values of the kinetic energy of the quintessence-like scalar field are negative, whereas those of the phantom-like one are positive in the neighbourhood of the bouncing point $ t \approx 3.93 $ in both $ \mathcal{F}(R,T) $ gravity ($ \lambda_1\neq 0 $) and GR ($ \lambda_2=0 $). 
The negative values of the kinetic energy of the quintessence-like scalar field  means that in the present dark energy model, the late-time cosmic acceleration can be realized by the repulsive force in the neighbourhood of the bouncing point $ t \approx 3.93 $ as found in Fig. 6(a). 
From Eq. (\ref{28}), it is seen that the potential energies $ V(\phi_q) $ and $ V(\phi_{ph}) $ for the quintessence and phantom-like scalar field are negligible in the neighbourhood of the bouncing point at $ t \approx 3.93 $ in $ \mathcal{F}(R,T) $ gravity as well as in general relativity. 
It is important to note that a scalar field with its negative potential energy and minimal coupling to gravity in general gravity can lead to the end of an accelerated contraction within a finite time and pull the trigger of the onset of the cosmic expansion in the vicinity of the bouncing point $ t \approx 3.93 $ as shown in Fig. 6(c). In addition, it is recognized that the sign of the kinetic energy of the quintessence-like and phantom-like scalar fields change in the neighbourhood of the bouncing point $ t \approx 3.93 $ only in $ \mathcal{F}(R,T) $ gravity, and such a behavior cannot be realized in general relativity, as represented in Fig. 6(a,b).

\section{Conclusions}
In the present paper, we have studied a non-singular bouncing behavior of the spatially flat homogeneous and isotropic space-time in $ \mathcal{F}(R,T) $ gravity with higher-order curvature terms. We have used the reconstruction technique by parametrizing the Hubble parameter $H$ in Eq. (\ref{15}). It has been found that the time when the cosmological bounce occurs (called the boucing point) delays as model parameters $\beta$ and $\gamma$ decrease as shown in Fig. 1. The possible range of $\beta$ and $\gamma$ for which the bouncing behavior can be realized from a contracting phase $ (H<0) $ to an expanding one $ (H>0) $ have been shown in Table I. It has been shown that an accelerated universe appears around the bouncing point $ t \approx 3.93 $ in the presence of dust or vacuum as depicted in Fig. 3. We have also examined the range of $\beta$ and $\gamma$ for which the accelerating criteria $ (q<0) $ is satisfied in Table II. 
 
Moreover, it has been demonstrated that the EoS parameter of the universe crosses the phantom divide line $ \omega=-1 $ in the neighbourhood of the bouncing point with satisfying the bouncing criteria as shown in Fig. 4(a). The realisation of the cosmological bounce is indicated by the time derivative of the Hubble parameter $ \dot{H}>0 $ and the violation of null energy condition (NEC) in the neighbourhood of the bouncing point $ t \approx 3.93 $ as shown in Fig. 2(a) and 4(b), respectively. We have discussed the quintessence-like and phantom-like scalar fields in $ \mathcal{F}(R,T) $ gravity with higher-order curvature terms and general relativity. The necessary condition in Eq. (\ref{30}) for the bouncing behavior is satisfied. 

The kinetic energy of a quintessence and phantom-like scalar field lead to a similar behavior of the universe to that in $ \mathcal{F}(R,T) $ gravity with higher-order curvature terms and general relativity owing to the differing scales as shown in Fig. 6(a) and Fig. 6(b). However, as depicted in Fig. 6(c), the potential energy of a quintessence and phantom-like scalar field is negative in $ \mathcal{F}(R,T) $ gravity with higher-order curvature terms and positive in general relativity. Hence, it can be interpreted that the canonical scalar field with a negative potential energy density and minimally coupled to the Einstein gravity induces the end of the accelerated contraction within a finite cosmic time and leads to the cosmic expansion in the vicinity of the bouncing point $ t \approx 3.93 $. In such a situation, for the present scenario, the universe behaves similarly to that in a quintom model with the requirement specified in Eq. (\ref{30}), which is an essential condition for the bouncing phenomenon. This condition is also supported in Refs. \cite{Cai:2007qw, Arefeva:2007wvo}. Thus, it has been concluded that the issue of the big bang singularity can be avoided in $ \mathcal{F}(R,T) $ gravity with higher-order curvature terms.

It is considered that the most crucial issue is whether the NEC-violating fields exist in nature. It may be observed at some point in future that the universe went through the bounce or Genesis epoch, and then it can be indicated that the NEC-violation indeed took place in the past \cite{Rubakov:2014jja,Creminelli:2006xe}.

\vskip0.2in 

\section*{Acknowledgments}
The work of KB was supported in part by the JSPS KAKENHI Grant Number JP21K03547. The authors express their sincere gratitude to the respected referee for his valuable comments and suggestions. The authors extend their sincere thanks to Prof. Sushant G. Ghosh from CTP, Jamia Millia Islamia, New Delhi, India, for engaging in insightful discussions, providing helpful suggestions, and editing the entire manuscript.

\vskip0.2in 

\appendix
\section{Detailed equations}

\begin{multline}\label{26}
\dot\phi^2_q=\frac{1}{2 \lambda _2 (\lambda _2+4 \pi ) t^3}6 \lambda _1 (\gamma  t^n (\lambda _2 (n (n^2+22 \alpha  (n-1) t-3 n+58 \alpha ^2 t^2+2)+2 \beta  t^2 (58 \alpha  (n+1) t+ \\
n (11 n-14))+58 \beta ^2 (n+2) t^4)+48 \pi  t (3 \beta ^2 (n+2) t^3+\beta  t (6 \alpha  (n+1) t+(n-2) n)+\alpha  n (n+3 \alpha  t-1)))+ \\
\gamma ^2 t^{2 n+1} (\lambda _2 (58 \beta  (2 n+1) t^2+n (19 n+116 \alpha  t-22))+24 \pi  (6 \beta  (2 n+1) t^2+n (n+12 \alpha  t-2)))+ \\
2 \gamma ^3 (29 \lambda _2+72 \pi ) n t^{3 n+2}+\beta  t^3 (\lambda _2 (58 (\alpha +\beta  t)^2-3 \beta)+24 \pi  (6 (\alpha +\beta  t)^2-\beta )))+ \\
t^2 (-\lambda _2 (\gamma  t^n (n+18 t (\alpha +\beta  t))+9 \gamma ^2 t^{2 n+1}+t (\beta +9 (\alpha +\beta  t)^2))-24 \pi  t (\alpha +\gamma  t^n+\beta  t)^2),
\end{multline}

\begin{multline}\label{27}
  \dot\phi^2_{ph}= \frac{1}{2 \lambda _2 (\lambda _2+4 \pi ) t^3}t^2(\lambda _2 (\gamma  t^n (n+18 t (\alpha +\beta  t))+9 \gamma ^2 t^{2 n+1}+t (\beta +9 (\alpha +\beta  t)^2))+
  \\24 \pi  t (\alpha +\gamma  t^n+\beta  t)^2)-6 \lambda _1 (\lambda _2 (\gamma  t^n (n (n^2+22 \alpha  (n-1) t-3 n+58 \alpha ^2 t^2+2)+
  \\2 \beta  t^2 (58 \alpha  (n+1) t+n (11 n-14))+58 \beta ^2 (n+2) t^4)+\gamma ^2 t^{2 n+1} (58 \beta  (2 n+1) t^2+n (19 n+116 \alpha  t-22))+
  \\58 \gamma ^3 n t^{3 n+2}+\beta  t^3 (58 (\alpha +\beta  t)^2-3 \beta ))+24 \pi  t (2 \gamma  t^n (3 \beta ^2 (n+2) t^3+\beta  t (6 \alpha  (n+1) t+(n-2) n)+
  \\\alpha  n (n+3 \alpha  t-1))+\gamma ^2 t^{2 n} (6 \beta  (2 n+1) t^2+n (n+12 \alpha  t-2))+6 \gamma ^3 n t^{3 n+1}+\beta  t^2 (6 (\alpha +\beta  t)^2-\beta ))),
\end{multline}

\begin{multline}\label{28}
    V(\phi_q)=V(\phi_{ph})=\frac{1}{4 \lambda _2 (\lambda _2+2 \pi ) t^3}6 \lambda _1 (-\gamma  t^n (\lambda _2 (n (n^2+28 \alpha  (n-1) t-3 n+76 \alpha ^2 t^2+2)+
    \\4 \beta  t^2 (38 \alpha  (n+1) t+n (7 n-10))+76 \beta ^2 (n+2) t^4)+48 \pi  t (3 \beta ^2 (n+2) t^3+\beta  t (6 \alpha  (n+1) t+(n-2) n)+
    \\\alpha  n (n+3 \alpha  t-1)))-2 \gamma ^2 t^{2 n+1} (\lambda _2 (38 \beta  (2 n+1) t^2+n (11 n+76 \alpha  t-14))+12 \pi  (6 \beta  (2 n+1) t^2+
    \\n (n+12 \alpha  t-2)))-4 \gamma ^3 (19 \lambda _2+36 \pi ) n t^{3 n+2}+2 \beta  t^3 (-\lambda _2 (38 (\alpha +\beta  t)^2-3 \beta )-12 \pi  (6 (\alpha +\beta  t)^2-\beta )))+
    \\t^2 (\lambda _2 (\gamma  t^n (n+24 t (\alpha +\beta  t))+12 \gamma ^2 t^{2 n+1}+t (\beta +12 (\alpha +\beta  t)^2))+24 \pi  t (\alpha +\gamma  t^n+\beta  t)^2).
\end{multline}

\end{document}